\documentstyle[prl,aps,epsfig,twocolumn]{revtex}

\begin{document}
\draft		
\wideabs{	
\title{Characteristic Relations of Type-I Intermittency in the Presence of Noise}
\author{Won-Ho Kye$^\dagger$ and Chil-Min Kim$^\ddagger$}
\address{National Creative Research Initiative Center for Controlling Optical Chaos,\\
Pai-Chai University, Taejon 302-735, Korea}
\maketitle

\begin{abstract}
Near the point of tangent bifurcation,
the scaling properties of the laminar length of type-I intermittency
are investigated in the presence of noise. 
Based on analytic and numerical studies, 
we show that the scaling relation of the laminar length is dramatically 
deformed from $\frac{1}{\sqrt{\epsilon}}$ for $\epsilon >0$ to 
$\exp\{\frac{1}{D}|\epsilon|^{3/2}\}$ for $\epsilon<0$ as $\epsilon$ passes 
the bifurcation point $(\epsilon=0)$. 
The results explain why two coupled R\"ossler oscillators exhibit deformation 
of the scaling relation of the synchronous length 
in the nearly synchronous regime.
\end{abstract}
}

\narrowtext

Intermittency is the occurrence of 
a randomly alternating signal
between long regular (laminar) phases and relatively short 
irregular bursts\cite{Ott}.
It is considered to be important as one of the 
routes to chaos in nonlinear dynamics.
There have been extensive studies to manifest 
the route in terms of experiment 
as well as theory\cite{Ott,PM,Eckman,Hirsch1,Hirsch2,Hu,Crutchfield,Kim}.
The scaling properties of the laminar length were studied for
the first time by Pomeau and Manneville in the Lorenz model\cite{PM}.
Based on the renormalization group equation [RGE] some other 
authors also investigated them\cite{Hirsch2,Hu,Crutchfield}.
Recently, it was reported that the reinjection mechanism 
is another important factor that dictates the scaling relation
of the laminar length\cite{Crutchfield,Kim}.

Because noise is not avoidable in real 
environments, consideration of it is important 
to studying the realistic properties of 
nonlinear dynamical system\cite{Gardiner}. 
The characteristics of 
nonlinear dynamical system in the presence of noise have been 
investigated by several authors based on the Fokker-Planck 
Equation [FPE]\cite{Eckman,Hirsch1,Crutchfield} and RGE\cite{Hirsch2,Hu},
since the studies of Brownian motion initiated the
stochastic modeling of the natural phenomena\cite{Gardiner}. 

The system without noise converges to fixed points when bifurcation
occurs but it does not exhibit infinite laminar phase 
under the succeeding random perturbation.  
So there is the possibility that 
the scaling properties show quite different features from those of the 
conventional ones. In this respect, the recent 
investigation \cite{Lee} comes into our notice 
which observed that the scaling properties of the laminar length
are deformed in the nearly synchronous regime of 
two coupled R\"{o}ssler oscillators \cite{Pikovsky1}.

In this Letter, we investigate type-I intermittency in the presence
of noise before and after the tangent bifurcation.
Solving the Fokker-Planck equation [FPE]\cite{Hirsch1,Hirsch2,Gardiner}, 
we derive the  scaling relations of
the laminar length in the closed channel (i.e., $\epsilon <0$) 
and present the results by using numerical solution and simulation.
Based on them, we also explain the scaling relations that 
grow exponentially in the nearly synchronous regime in 
two coupled R\"ossler oscillators and eyelet intermittency\cite{Lee,Grebogi}.

The local Poincar\'e map of type-I intermittency 
in the presence of noise is described as the 
following difference equation
\cite{Ott,PM,Hirsch1,Hirsch2,Hu,Crutchfield,Kim},
\begin{equation}
x_{n+1}=x_n + a x_n^2 + \epsilon +\sqrt{2D} \xi_n,
\end{equation}
where $a$ is the positive arbitrary constant, $\epsilon$ 
the channel width between diagonal and map, 
and $D$ the dispersion of Gaussian noise $\xi_n$.
In the long laminar region, we can approximate the difference equation to the
stochastic differential equation as follows\cite{Hirsch1}:
\begin{equation}
\dot{x}= -V^\prime (x) +\sqrt{2D} \xi(t),
\end{equation}
where dot and prime denote the differentiation with respect 
to $t$ and $x$, respectively, $\xi(t)$ is the Gaussian white
noise such that $\langle \xi(t^\prime)\xi(t)\rangle=\delta(t^\prime-t)$ and 
$\langle \xi(t) \rangle=0$\cite{Gardiner}, and $V(x)$ is the potential given by 
$V(x)=-\frac{1}{3} a x^3 -\epsilon x + c$ where $c$ is the integration constant.
The above equation can be considered as the equation of motion of the
point particle under the potential $V(x)$ and random perturbation $\xi(t)$.
The relation between return map and potential is given in Fig. 1. In this
figure the stable and the unstable fixed points correspond to 
the extremal points of the potential.

\begin{figure}
\begin{center}
\rotatebox[origin=c]{0}{\includegraphics[height=4.5cm, width=5.0cm]{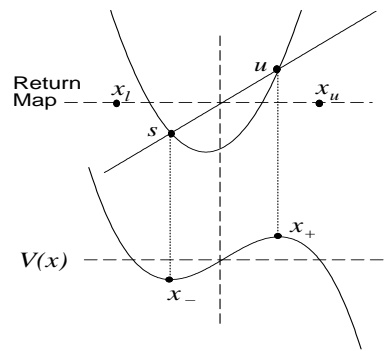}}
\end{center}
\caption{The relation between fixed and extremal points for $\epsilon<0$. 
$s$ and $u$ are the stable and unstable fixed points, $x_l$ and $x_u$ are
the lower and upper bounds of laminar phase, respectively 
and $x_-$ and $x_+$ are the extremal points of the potential.} 
\end{figure}

From the above stochastic differential equation,
we obtain the backward FPE\cite{Hirsch1,Gardiner} by following the
well-established procedure \cite{Gardiner} as follows:
\begin{equation}
\frac{\partial G(x, t)}{\partial t} = -V^\prime (x) \frac{ \partial G(x,t)}{\partial x}
+D \frac{\partial^2 G(x,t)}{\partial x^2},
\end{equation}
where $G(x, t)$ is the probability density of particle at $\{x, t\}$.
We obtain a mean first-passage time [MFPT] equation 
after integrating the above FPE with respect to time 
as follows\cite{Hirsch1,Gardiner}:
\begin{equation}
	-1=-V^\prime(x) \frac{d T}{d x} + D \frac{d^2T}{dx^2},
\end{equation}
where $T(x)$ is the mean escaping time defined by 
$T(x)=\langle t \rangle=-\int ^{\infty}_{0} t \frac{\partial G(x,t)}{\partial t} dt$ 
under the boundary conditions that $G(x, 0)=1$ and $\lim_{t \rightarrow \infty} G(x, t)=0$.
The MFPT function $T(x)$ is 
the average transition time
from the reinjection to the escaping point of the particle 
under the potential $V(x)$ and random perturbation. 

The general solution of Eq. (3) can be derived as follows:
\begin{eqnarray}
T(x) &=&c \int^x_{x_{l}} d x^{\prime} \exp\left\{\frac{1}{D}V(x^\prime)\right\} \nonumber\\
 & &-\frac{1}{D} \int^x_{x_{l}} dx^{\prime} \int_{x_{l}}^{x^{\prime}} dx^{\prime\prime}
\exp\left\{\frac{1}{D}(V(x^{\prime})-V(x^{\prime\prime}))\right\}, 
\end{eqnarray}

where $c$ is the integration constant, $x_{l}$ 
is the lower bound of the laminar phase, and $x$ is 
the destination point of the transition.
We can easily verify that Eq. (5) is the general solution 
for the MFPT equation by inserting Eq. (5) into Eq. (4). 

If noise is small enough such that $D \ll 1$, 
the first term in the above equation is suppressed by the factor of $1/D$
and the second term becomes dominant.
The second term is not integrable analytically. 
Then  we can expand the potential at the extremal point $x_\pm$ approximately
(see Fig. 1) such that
$V(x)\approx V(x_\pm)+\frac{V^{\prime \prime}(x_\pm)}{2}(x-x_\pm)^2+O((x-x_\pm)^3)$.

In that case, the MFPT function $T(x)$ can be approximated as follows: 
\begin{eqnarray}
T(x) &\approx& -\frac{1}{D}\exp\left\{\frac{1}{D}
(V(x_+)-V(x_-))\right\} \times \nonumber\\
& &\int^{x}_{x_l}dx^\prime \int^{x^\prime}_{x_l} dx^{\prime \prime} 
\exp\left \{\frac{1}{2D}
\left[ V^{\prime\prime}(x_+)(x^{\prime}-x_+)^{2} \right.\right. \nonumber\\
& &\left.\left.-V^{\prime\prime}(x_-)(x^{\prime\prime}-x_-)^{2}\right]\right\}.
\end{eqnarray}
The extremal points are given by $x_\pm=\pm \sqrt{-\epsilon/a}$ in Eq. (2). 
In the far outside of the laminar phase (i.e., 
at the limit $x\rightarrow \infty$ and  $x_l \ll x_{-}$), 
we can perform the integration of the quadratic exponent \cite{Gardiner}
and then obtain the following 
approximated solution of the MFPT equation:
\begin{equation}
|T| = \frac{\pi}{\sqrt{a |\epsilon|}} \exp 
\left \{\frac{4}{3D\sqrt{a}} |\epsilon|^{3/2} \right\}\mbox{~~~for~~}\epsilon <0.
\end{equation}  
The above solution is consistent with the formal one which was derived
in the previous investigation by the FPE and RGE analysis 
\cite{Eckman,Hirsch1,Hirsch2,Crutchfield} such that 
$\langle l \rangle\sim \epsilon^{-1/2}f(\sigma^2/\epsilon^{3/2})$.
We remark the fact that there has been no explicit derivation
in analytic form like Eq. (7). The analytic solution is important
in analyzing  various intermittent phenomena quantitatively.
In particular we are interested in the mysterious deformation 
of the type-I scaling near phase synchronization regime \cite{Lee}
and eventually show that it enables the theoretical understanding
of those phenomena.

After taking the logarithm on Eq. (7), we obtain the equation such that
$\ln T \sim -\frac{1}{2} \ln |\epsilon| +\frac{1}{D} |\epsilon|^{3/2}$.   
Our main interests are  in the far region ($|\epsilon| \gg 0$) 
from the bifurcation point 
because the transition from the intermittency to stable orbit 
occurs here.
In this region the scaling is dominated by the second term such that 
$\ln T \sim \frac{1}{D} |\epsilon|^{3/2}$ 
(note that this exponential saturation rapidly forms from  
$|\epsilon| \geq 1.0 \times 10^{-3} $ because the noise 
is small enough such that $\frac{1}{D}=0.5\times 10 ^{+6}$ in our simulation  
(see Fig. 3 (b)) ).
 
The reinjection probability $P(x_{in})$ was reported to be another important factor
which affects the scaling relations of the laminar length\cite{Kim}. 
And it was generally considered to obtain average laminar length\cite{Crutchfield,Kim}. 
But in this investigation we only consider the fixed reinjection probability 
$P(x_{in})=\delta(x_{in}-\Delta)$ to study the intrinsic scaling property 
of the system (note in all of our simulation we set the 
reinjection point $\Delta=x_l$).

Note that the approximation procedure used in Eq. (6) 
is not applicable for $\epsilon >0$ and a transient region
because the equation is a solution for $\epsilon <0$ where the potential 
has extremal points (see Fig. 1). 
At the far outside from the bifurcation point (i.e., $\epsilon \gg D >0$),
we take the limit D $\rightarrow 0$ in Eq. (4) and obtain the
conventional scaling relation of type-I intermittency
$T \sim \frac{1}{\sqrt{\epsilon}}$\cite{PM,Hirsch1}.
For intermediate range $\epsilon \sim D$, we present 
the numerical results of the scaling relation of the laminar length
in Fig. 3 (c) and (d).

In Eq. (7), if we take $ {D \rightarrow 0}$ the MFPT function $T \rightarrow \infty$.
So the particle trapped in the well does not escape from it
if the random perturbation is turned off. 

\begin{figure}
\begin{center}
\rotatebox[origin=c]{0}{\includegraphics[height=4.0cm, width=8.5cm]{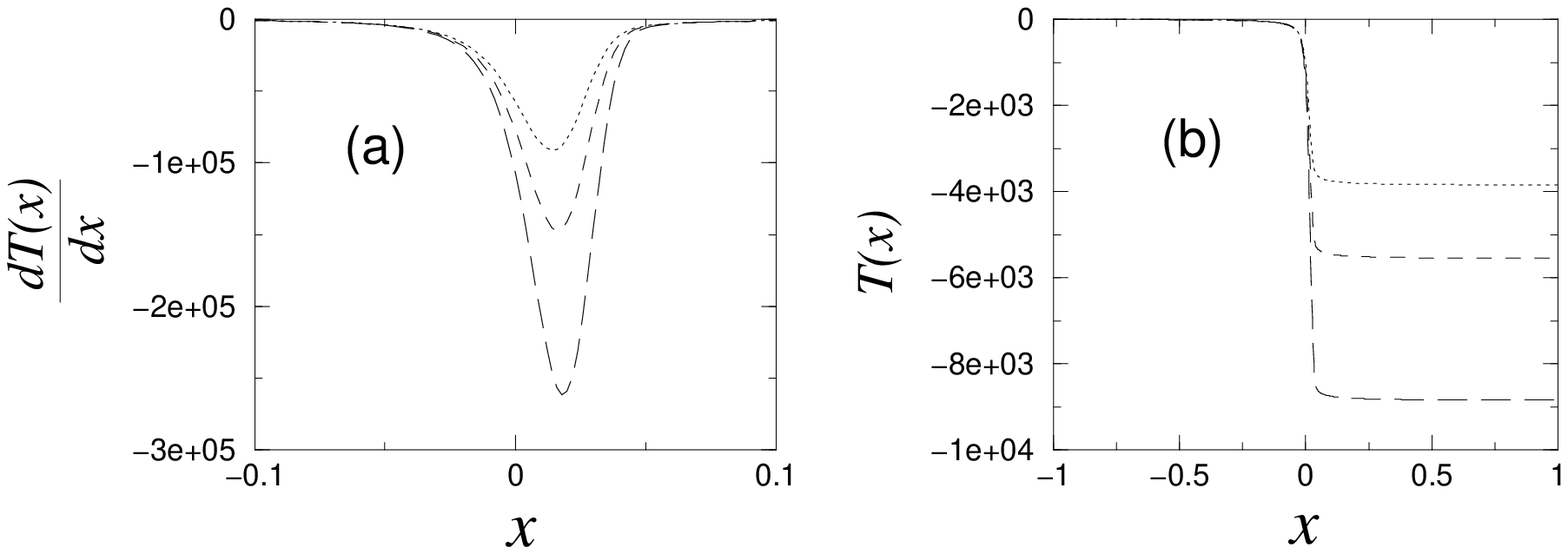}}
\end{center}
\caption{The numerical solutions of the MFPT equation.
(a) the first differentiation of the MFPT function and (b) 
the MFPT function when $a=0.1$ and $D=2.0\times 10^{-6}$.
The dotted, dashed, and long dashed lines are for 
$\epsilon=-1.0\times 10^{-5},$ $\epsilon=-2.0\times 10^{-5},$ and
$\epsilon=-3.0\times 10^{-5},$ respectively.}
\end{figure}

The scaling relation of the laminar length can be
verified in numerical solutions and simulations.
In Fig. 2, the numerically solved the MFPT function $T(x)$ shows the 
typical kink shape\cite{Rajaraman},
thus we can know that $T(x)$ is the good physical quantity reflecting 
the transition characteristics of intermittency from the laminar phase
to chaotic burst. 
In that case we can define a topological index of the transition such that 
$Q = |T(\infty)-T(-\infty)|$\cite{Rajaraman}.
The negative signature in Fig. 1 stems from the backwarding property of
FPE of Eq. (3) and the absolute value of the MFPT function $T(x)$ is the 
laminar length\cite{Hirsch1,Gardiner}.

In the following presentation of the numerical results,
we let $x_{l} =-1.0$ and $x_{u}=1.0$ 
and the average laminar length 
is $ \langle l \rangle \equiv |T(x_u)-T(x_l)|$ (as given in Fig. 2 (b), 
$T(x)$ rapidly converges to the constants
outside the center, so that we can say that $\langle l \rangle$ is
a kind of topological index as defined above). 
To confirm the scaling relation of Eq. (7), we not only perform a direct 
simulation with Eq. (1) but also solve 
the second order MFPT equation (Eq. (4)) numerically. 
\begin{figure}
\begin{center}
\rotatebox[origin=c]{0}{\includegraphics[height=8.5cm, width=8.5cm]{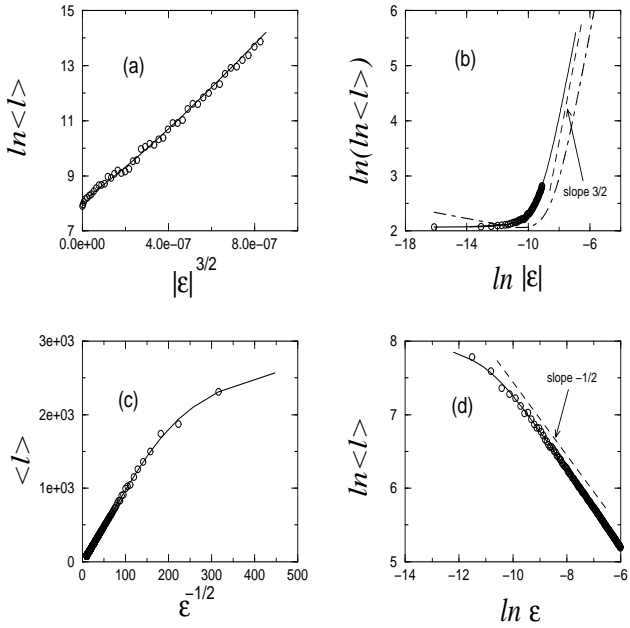}}
\end{center}
\caption{The laminar scaling before and after the tangent bifurcation. 
(a) and (b) are for $\epsilon <0$. (c) and (d) are for $\epsilon >0$.
The circles are simulation data from Eq. (1) and the solid 
lines are solution data from Eq. (4). 
The dashed lines in (b) and (d) show the  slope 3/2 and -1/2 
saturation, respectively and the dot-dashed line in (b) is 
an analytic solution curve (Eq. (7)) 
(the maximum and dispersion of the Gaussian 
noise are $|\xi|=2.0\times 10^{-3}$ and $D=2.0\times 10^{-6}$, respectively).}
\end{figure}
In Fig. 3, the circle points and solid lines are simulation data from Eq. (1)
and solution data from Eq. (4), respectively.
The dashed lines in Fig. 3 (b) and (d) are $3/2$ scaling expected from
Eq. (7) and $-1/2$ scaling for $\epsilon >0$, respectively.
When we simulate Eq. (1), the approximated Gaussian noise is used
(see the caption of Fig. 3 for details). The solution data 
agree well with the simulation ones.  We note here that there are 
some shifts from the solution line when the uniform noise is applied,
but the scaling behaviors are invariant in both cases.

Fig. 3 (a) shows the scaling relation for $\epsilon <0$. The figure,
$\ln \langle l \rangle $ as a function of $ |\epsilon|^{3/2}$ shows 
a straight line approximately
to confirm $\langle l \rangle \sim \exp\{\frac{1}{D}|\epsilon|^{3/2}\}$.
The exponent $3/2$ appears more clearly when we obtain $\ln(\ln \langle l \rangle)$
as a function of $\ln(\epsilon)$ as given in Fig. 3 (b).
Thus we can verify the slope is eventually saturated to $3/2$.
This is the very scaling behavior obtained in Eq. (7) analytically.
The analytic solution is also given in Fig. 3 (b) as the dot-dashed line
that is the plotting of Eq. (7).
In figures though we present the simulation data within the limit 
of numerical calculation of the map (Eq. (1))
the data well follows the deformation of the scaling behavior.

In Fig. 3 (c), the conventional scaling 
$\langle l \rangle\sim \frac{1}{\sqrt{\epsilon}}$
holds for relatively wide channel region ($\epsilon \gg D >0$)
so that the slope -1/2 saturation can be obtained in Fig. 3 (d). 
As the channel
width $\epsilon$ becomes close to zero, the straight line begins to 
bend (Fig. 3 (c)) and  after $\epsilon$ passes zero point, 
the straight line reappears (Fig. 3 (a)).

We now apply this analysis to two coupled R\"{o}ssler oscillators
which are a good laboratory for the studying of 
type-I intermittency with random perturbation\cite{Lee,Pikovsky1,Pikovsky2}. 
It is important to discuss the correspondence between two coupled 
R\"{o}ssler oscillators and type-I intermittency in the presence of noise
to explore the origin of nearly synchronous phenomena.
The two coupled R\"{o}ssler oscillators are given as 
follows\cite{Lee,Pikovsky1,Pikovsky2}:
\begin{eqnarray}
\dot{x}_{1,2}&=&-\omega_{1,2} y_{1,2} -z_{1,2} +\epsilon(x_{2,1}-x_{1,2}),\nonumber\\
\dot{y}_{1,2}&=& \omega_{1,2} x_{1,2} +0.15 y_{1,2},\\
\dot{z}_{1,2}&=& 0.2 +z_{1.2}(x_{1,2} -10.0), \nonumber
\end{eqnarray}
where $\omega_{1,2}=1.0\pm 0.015$.
The phase difference between the two oscillators can be rewritten as follows: 
\begin{equation}
\frac{d}{dt}(\theta)= F(\theta, \epsilon)+G(\phi_1, \phi_2),
\end{equation}
where,
\begin{eqnarray*}
F(\theta,\epsilon)&=&\omega_1-\omega_2-\frac{\epsilon}{2}
	 [\frac{A_2}{A_1}+\frac{A_1}{A_2}]\sin\theta,\\
G(\phi_1,\phi_2)&=& 0.15 (\sin \phi_1 \cos\phi_1 -\sin \phi_2 \cos\phi_2)\\
		& &+(\frac{z_1}{A_1} \sin\phi_1 -\frac{z_2}{A_2} \sin\phi_2).
\end{eqnarray*}
where $\theta =\phi_1-\phi_2$, $A_{1,2}=\sqrt{x_{1,2}^2 + y_{1,2}^2}$
and $\phi_{1,2}=\arctan(y_{1,2}/x_{1,2})$.
In the above equations we neglect the fast fluctuation term 
which depends on $\epsilon$, 
in $F(\theta, \epsilon)$ as discussed in Ref. \cite{Lee}.

As already discussed in Ref. \cite{Lee}, the potential of this system 
$V(\theta)= -\int d\theta F(\theta, \epsilon)$ 
shows the saddle node bifurcation at 
$\epsilon=\epsilon_t(=0.0276)$.
If we identify $G(\phi_1, \phi_2)$ as the random
perturbation term of  Eq. (1),
we can argue the scaling relation of the length of synchronization in
nearly synchronous regime of these two coupled oscillators are effectively 
similar to that of type-I intermittency with noise (Eq. (2))
(note: the dispersion of $G(\phi_1, \phi_2)$ hardly
depends on $\epsilon$ for $\epsilon_t < \epsilon < \epsilon_c$). 
Thus we remark the scaling of the laminar length is separated into two regions
in the center of $\epsilon=\epsilon_t$.
So it has
$\langle l \rangle \sim \exp\{|\epsilon_t-\epsilon|^{3/2}\}$ 
for ${\epsilon > \epsilon_t}$ and 
$\langle l\rangle=|\epsilon_t-\epsilon|^{-1/2}$ 
for ${\epsilon< \epsilon_t}$ like the previously presented results 
(note: based on numerical studies, Ref. \cite{Lee} proposed the scaling 
$\langle l \rangle\sim \exp\{-|\epsilon_c-\epsilon|^{1/2}\}$
for $\epsilon > \epsilon_t (\epsilon_c=0.0286)$).

As in the above analysis, 
the persistency of the intermittency is caused by 
the random perturbation $\xi(t)$ in closed channel region. 
Thus it can be argued that the nontrivial $2\pi$ phase
jumps observed in Ref. \cite{Lee} originates from the random
perturbation term $G(\phi_1, \phi_2)$, neglected in the discussion 
of Ref.\cite{Lee}, rather than from the term $\frac{A_2}{A_1} + \frac{A_1}{A_2}$.
We perform the numerical simulation for the scaling of the laminar length 
in both regions ($\epsilon < \epsilon_t$ and $\epsilon > \epsilon_t$),  
and the results are presented in the Fig. 4. 
As we expected, the figures well agree with 
the previous theoretical analysis of the scaling relation.

\begin{figure}
\begin{center}
\rotatebox[origin=c]{0}{\includegraphics[height=7.0cm, width=5cm]{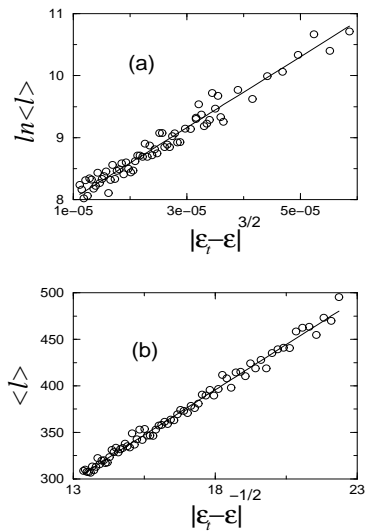}}
\end{center}
\caption{The laminar scaling in the two coupled R\"ossler oscillators for
(a) $\epsilon > \epsilon_t$ and (b) $\epsilon < \epsilon_t$.}
\end{figure}

In conclusion, the new scaling relation of type-I intermittency
is presented in the region after the tangent bifurcation.
We observe that the conventional scaling
relation $\langle l \rangle \sim\frac{1}{\sqrt{\epsilon}}$ of type-I intermittency
holds only in relatively wide channel ($\epsilon \gg D >0$) and 
it begins to deform as $\epsilon$ approaches zero.
The scaling relation is eventually saturated by
$\langle l\rangle \sim \exp\{\frac{1}{D}|\epsilon|^{3/2}\}$ (Eq. (7) and Fig 3.  (a) and (b))
after the tangent bifurcation ($\epsilon <0$). 
Such dramatic deformation of the scaling relation stems from the persistency
of the intermittency with the random perturbation even though the system is 
in a state of closed channel region (i.e., $\epsilon < 0$).
This is why the laminar length in closed channel region 
grows faster than that of the positive one\cite{Pikovsky2,Grebogi}.
From these results, we can also explain 
why two coupled R\"ossler oscillators exhibit deformation 
of the scaling relation of the synchronous length 
in the nearly synchronous regime\cite{Lee}.

The authors thank J. P. Eckmann and P. Wittwer for helpful comments 
and S. Rim, D.-U. Hwang, and I. B. Kim for valuable discussions. 
This work is supported by Creative Research Initiatives of the
Korea Ministry of Science and Technology.

\end{document}